\documentstyle[psfig,epsf]{article}
\begin{document}

\begin{titlepage}

\title{\begin{flushright}
{\normalsize FTUAM 96/2}
\end{flushright}
\vspace{2cm}
{\Large \bf Approximate solutions in scalar and fermionic theories within
the exact renormalization group approach \thanks{Talk given by Yu.K.
at the Xth Workshop on High Energy Physics and Quantum Field Theory,
Zvenigorod (Russia), 20-26 September, 1995}}}

\author{Jordi Comellas \thanks{E-mail: comellas@sophia.ecm.ub.es}\\
Departament d'Estructura i Constituents de la Mat\`eria, 
Facultat de F\'\i sica \\ 
Universitat de Barcelona, Diagonal~647,~08028 Barcelona, Spain\\
Yuri Kubyshin \thanks{On leave of absence from the Institute of Nuclear
Physics, Moscow State University, 119899 Moscow, Russia.}
\thanks{E-mail: kubyshin@delta.ft.uam.es}\\
Departamento de F\'\i sica Te\'orica C-XI, Universidad Aut\'onoma de
Madrid \\
Cantoblanco, 28049 Madrid, Spain \\
and \\
Enrique Moreno\thanks{E-mail: moreno@scisun.sci.ccny.cuny.edu}\\
Department of Physics, City College of New York\\
New York, NY 10031, U.S.A.}

\date{January 18, 1996}

\maketitle

\begin{abstract}

We give a review of the exact renormalization group (ERG) approach 
and illustrate its applications in scalar and fermionic theories.  
The derivative expansion and approximations based on the derivative 
expansion with further truncation in the number of fields (mixed 
approximation) are discussed. We analyse the mixed approximation 
for a three-dimensional scalar theory and show that it is less 
effective than the pure derivative expansion. For pure fermionic 
theories analytical solutions for the pure derivative expansion 
and mixed approximation in the limit $N \to \infty $, where $N$ 
is the number of fermionic species, are found. For
finite $N$ a few series of fixed point solutions with their anomalous
dimensions and critical exponents are computed numerically.  We argue
that one of the fermionic solutions can be identified with that of
Dashen and Frishman, whereas the others seem to be new ones. The issues
of spurious solutions and scheme dependence of the results are
discussed.

\end{abstract}
\end{titlepage}

\section{Introduction}

One of the methods, which, in principle, is capable of handling
non-perturbatively many problems of quantum field theory (such as
calculation of exact renormalization flows, explicit derivation of
low-energy effective Lagrangians from high-energy ones, extension of the
c-theorem to higher dimensions, etc.) is the exact renormalization group
(ERG, hereafter).  Originally developed by Wilson in his seminal
articles in the early seventies \cite{Wil} (see Ref.~\cite{WK} for a
classical review), it has recently attracted much attention.

In this article we are going to consider a version of the ERG approach
based on a functional differential equation.  It determines how the
action changes as momentum modes are integrated out in lowering down a
characteristic momentum scale (cutoff) $\Lambda$ while keeping the
$S$-matrix elements invariant.  This is the so-called ERG equation.
Using it for the most general action, consistent with the symmetries of
the model under consideration, the complete set of $\beta $-functions
(or flow equations) can be derived, and from them the location of fixed
points, critical exponents, behaviour of the renormalization flows,
low-energy effective action, etc. can be computed in principle.
However, with known techniques it is not possible to handle such general
actions for physically interesting models.  One needs to choose an
approximation which, on one hand, reduces sufficiently the complexity of
the problem and, on the other hand, captures the essential features of
the physical system under consideration.  Approximations most commonly
used in the literature are the local potential approximation \cite{HH} -
\cite{Maggiore}, polynomial approximation for the effective potential,
which can be regarded as further simplification of the local potential
approximation, \cite{Margaritis}, \cite{HKLM} and the derivative
expansion \cite{der} - \cite{BHLM}. Discussion of their main features
and some results obtained by these techniques are given in Sect.~2.3.
For calculations we will use the derivative expansion in the case of
fermions and the derivative expansion combined with expansion in powers
of fields (mixed approximation) for fermionic and scalar theories.

In this article we will consider the Polchinski type ERG equation
\cite{Pol} with its derivation being discussed along the lines of 
Ref.~\cite{Ball}.  
It is worth mentioning that there are other types of ERG
equations used in quantum field theory calculations.  Among them is the
one of Ref.~\cite{HH}, based on the sharp cutoff realization of the
Wilson idea by Wegner and Houghton \cite{WH}, the average effective
action equation \cite{AEA}, a version of the ERG for the effective
action with a sharp cutoff of Ref.~\cite{Morrisiii}, to name a few.

Although firstly used for studies of critical phenomena in condensed
matter problems \cite{Fisher}, now the scope of the ERG approach covers 
many other areas \cite{Vvedensky}, 
including field theory, as it was demonstrated by
Polchinski in an elegant paper where he proved the perturbative
renormalizability of $\lambda \phi^4$ in a quite simple way \cite{Pol}.
Most of the work was devoted to scalar theories, both to their general
studies \cite{Pol}, \cite{scalargen} and to computation of fixed points
and the critical exponents \cite{HH} - \cite{Morrisi}, \cite{Margaritis}
- \cite{HKLM}, \cite{Morrisiia} - \cite{Morrisiii}.  Much work has been
done on gauge theories, see Refs.~\cite{gauge} for examples.  However,
a satisfactory formulation of the approach is lacking in this case, the
main difficulty being to maintain the gauge invariance along the
renormalization flow of the regularized action.  It is worth mentioning
studies of phase transitions at finite temperature \cite{fin}, of bound
states \cite{bound}, \cite{UW} and in relativistic cosmology
\cite{cosmology} within the ERG approach.  Investigations of theories
with fermions, which have been done so far, are very restricted.  In
\cite{fermioni} the ERG approach was used for proofs of the
renormalizability and convergence of the perturbative expansion of the
two-dimensional Gross-Neveu model. Discussions of physical effects in
many of the articles are limited to  perturbative expansions \cite{QED}
or models with fermions interacting
with scalars either through Yukawa coupling \cite{yukawa} or through
coupling of the fermion bilinears to a scalar potential and then
studying the fixed points and renormalization flows for this potential
\cite{Maggiore}, \cite{UW}, \cite{fermionii}.
The ERG formalism for pure fermionic theories, in which
fermionic degrees of freedom are treated directly and
non-perturbatively, has been developed recently in Ref.~\cite{CKM}.
Fixed point solutions and the critical exponents were also analysed
there.

The purpose of this article is to give an account of basic ideas of the
approach and main steps of the derivation of the ERG equation (for the
scalar and fermionic theories) and to describe main techniques used for
non-perturbative calculations of fixed points and critical exponents.
We will also give a review of main results of such calculations and
illustrate the application of the techniques in two concrete models.
One of them is a three-dimensional model of one-component scalar field
with $Z_{2}$-symmetry. It possesses the non-trivial fixed point (Wilson
fixed point), and was analysed by many authors (see \cite{HH},
\cite{Morrisi}, \cite{Margaritis} - \cite{HKLM}, \cite{Morrisiia} -
\cite{BHLM}).  The other example is a two-dimensional Gross-Neveu type
theory which seems to be a perfect site for such studies:  its phase
diagram is non-trivial and, on the other hand, the two-dimensional
Clifford algebra is the simplest one and generated by the Pauli
matrices.  In the analysis of this model we will follow mainly 
Ref.~\cite{CKM}, but also some new results on the pure derivative 
expansion will be presented.

We would like to mention that pure fermionic
theories, analysed in this article, are widely used in description of
fundamental interactions.  There are some interesting phenomenological
models of this kind, like the celebrated Fermi theory of weak
interactions \cite{Fermi}, models of resonance physics \cite{QCD} based
on extensions of the Nambu-Jona-Lasinio action \cite{NJL}, models
explaining the symmetry-breaking sector of the electroweak theory,
especially in connection with technicolor theories \cite{tech}, etc.
Also we would like to remind that fermions are not always easily
manageable by non-perturbative methods, e.g. in lattice formulation.  On
the contrary, as we will show, there are no principal difficulties in
treating fermions within the ERG approach.  Moreover, once truncated,
the ERG equations for fermions and bosons look very similarly, thus
making possible applications of techniques, already known in bosonic
theories, in the case of fermions.

Moreover, the two-dimensional Gross-Neveu model \cite{GN} , which
contains $N$ species of fermions transforming under a global
representation of the unitary group $U(N)$, is interesting in its own.
It is asymptotically free and renormalizable within perturbation theory
and also within the $1/N$ expansion.  However, none of these
approximations is capable to find any non-trivial fixed point for $d=2$.
An interesting modification leads to the so-called chiral Gross-Neveu
model \cite{GN}, which is chosen to have the additional symmetry of the
$U(1)$ chiral group.  As in the previous case no fixed points, besides
the Gaussian one, can be found within the $1/N$ approximation. However,
using non-perturbative methods based on current algebra and conformal
techniques Dashen and Frishman in Ref.~\cite{DF} found two critical
curves in the space of couplings for which the theory is scale
invariant.  One of the lines corresponds to the abelian Thirring model,
whereas the other one is truly non-trivial and does not pass through the
origin.  A very remarkable fact is that this result is exact and is not
given by a zero of a $\beta$-function, neither in the perturbation theory
nor in the large $N$ expansion. For this two continuum sets of critical
theories the value of the coupling constant  associated to the
$SU(N)$ degrees of freedom is fixed to be equal to zero or $4 \pi
/(N+1)$ respectively, while the coupling associated to the abelian
degrees of freedom is arbitrary. More recently, 
using bosonization, current algebra and
conformal techniques, other non-trivial fixed points in two-dimensional
fermionic models were found \cite{ofp}.

The article is organized as follows.  In Sect.~2 we derive the ERG
equation for scalar theories and study the mixed approximation for a
three-dimensional model.  Results of computation of the fixed points and
critical exponents are compared with those obtained by the pure
derivative expansion and the polynomial approximation in earlier works.
In Sect.~3 we derive the ERG equation for pure fermionic theories for
arbitrary number of spacetime dimensions.  Sect.~4 is devoted to the
construction of the action.  The chiral Gross-Neveu type model, we are
going to study, is defined through its symmetries, and the
approximations, we are going to use, are explained.  In Sect.~5 we work
out the pure derivative expansion and obtain analytically two fixed
point solutions.  The calculation of the $\beta$-functions, fixed points
and critical exponents within the mixed approximation in the fermionic
model is covered by Sect.~6. Sect.~7 contains discussion of the results
and some conclusions.

\section{ Scalar theory }

In this section we explain the main points of the derivation of the ERG
equation for the scalar case following the lines of Ref.~\cite{Ball}.
We also present an example of calculation of the fixed point and the
critical exponents in a scalar $Z_{2}$-symmetric model in three
(Euclidean) dimensions in order to illustrate how the whole machinery
works.

As usual in this type of problems, there are three important steps.  The
first one is to derive an ERG equation which governs the behavior of
the action as we integrate out modes, that is, how the action of our
effective theory has to be modified when we vary the characteristic
scale (cutoff) $\Lambda $. This is achieved by requiring the
independence of the S-matrix elements on $\Lambda $. However, it
is more convenient to impose a stronger condition, namely the
independence of Green's functions on $\Lambda$.  The second step is to
choose the appropriate ``space of local interactions''.  This amounts
to define properly the theory and a sensible
approximation of the general action.  The last step is to compute the
$\beta $-functions, which characterize the RG flow of the theory, and to
obtain physically relevant information from it.  We may be interested in
global properties of the flow, like the number of the fixed points it
contains, or just local ones, like the behavior of the flow in the
vicinity of them.  The latter is characterized by the critical
exponents, which are, moreover, universal quantities.

\subsection{ERG equation}

To derive an ERG equation we impose the condition that the generating
functional of Green's functions
\[
{\cal Z}[J] = \int D\phi \exp \left(-S\left[ \phi ;\Lambda \right]
+\int_pJ_pQ^{-1}_{\Lambda }\phi_{-p}+f_{\Lambda }\right)  
\]
is independent of some cutoff $\Lambda$ that sets the scale of the theory.
Here $J_p$ is an external source, $Q_{\Lambda}$ a regulating function and
$f_{\Lambda}$ is a {\it c}-number quantity.

The action is arbitrarily divided into a ``kinetic term" and an interaction
part as
\[
S\left[\phi;\Lambda\right]\equiv{1\over
2}\int_p\phi_p\phi_{-p} P^{-1}_{\Lambda}\left( p^2\right) +
S_{int}\left[ \phi ; \Lambda\right],      
\]
where $P_{\Lambda}$ stands for the regulated free propagator
\begin{equation}
P_{\Lambda}\left( p^2\right)
=\left( 2\pi \right)^d{K \left( {p^2 \over \Lambda^2} \right)
\over p^2}   \label{n3}
\end{equation}
with $K (z) $ being
an arbitrary (but fixed) cutoff function which vanishes faster than any
polynomial when $p^2 \to \infty$ and satisfies $K(0)=1$.

Using the path integral identity
\begin{equation}
\int D\phi \ {\delta \over \delta \phi_p} \left( {1\over 2}{\delta
\over \delta \phi_{-p}}+P^{-1}_\Lambda \phi_p-Q^{-1}_\Lambda
J_p\right) \ e^{-S\left[ \phi ;\Lambda \right]
+\int_pJ_pQ^{-1}_{\Lambda }\phi_{-p}+f_{\Lambda }}=0,  \label{n4}
\end{equation}
we can write the condition of the independence of the generating
functional on the cutoff
$\Lambda d{\cal Z}/d\Lambda=0$
as an evolution equation for the total action $S[\phi,\Lambda]$
\begin{equation}
\left<\dot{S}\right> = \left<{1\over 2}\int_p \dot{P}_{\Lambda}\left(
p^2\right) \left( {\delta S \over \delta \phi_p}{\delta
S\over \delta \phi_{-p}}-{\delta^2S\over \delta \phi_p\delta
\phi_{-p}}\right) - \int_p {\delta S \over \delta \phi_{p} }
 \dot{P}_{\Lambda} P^{-1}_{\Lambda} \phi_p\right>    \label{n5}
\end{equation}
provided
\[
f_{\Lambda }=
 \int dt \left\{ \int_p \tilde Q^{-2}\left( p^2\right)
 {\dot P_{\Lambda }\left( p^2\right) \over P_{\Lambda }^{2}
 \left( p^2\right)} J_p J_{-p} - \int_{p}
  {\dot P_{\Lambda }\left( p^2\right) \over P_{\Lambda }\left(
  p^2\right)} \delta (0) \right\} \ ,   
\]
where $Q_{\Lambda }\left( p^2 \right) =
P_{\Lambda }\left( p^2 \right) \tilde Q\left( p^2\right)$
and $\tilde Q\left(p^2\right) $ satisfies $\dot {\tilde
Q}\left( p^2\right) =(\eta/ 2) \tilde Q\left( p^2\right) $.
(We have defined ${\dot a} \equiv  d a/ d t$, with $t=-{\rm log}
(\Lambda/\Lambda_0)$ and $\Lambda_0$ being a fixed scale ).
In Eq (\ref{n5}) we have taken into account the non-trivial evolution of
the field itself: $\dot{\phi}_p\equiv (\eta / 2)\phi_p$ and the
derivatives with respect to $\Lambda$ should be understood as acting
only on the coefficients of the action, not on the fields.  This convention
will be used throughout this paper.

Note that even though Eq.~(\ref{n5}) depends on the source $J_p$ through the
Boltzmann weight, the argument of the v.e.v's does not. So Eq.~(\ref{n5})
is valid for arbitrary $J_p$ if and only if the arguments
of the r.h.s and l.h.s
vacuum expectation values are equal. So from now on  we will
drop the v.e.v brackets.

To complete the renormalization group transformation we should
add the canonical rescalings of the couplings. The easiest way to proceed
is to re-write Eq.~(\ref{n5}) in
terms of dimensionless quantities by extracting the appropriate powers of
$\Lambda$ carrying the canonical scale dependencies. We get
\begin{eqnarray}
\dot{S} & = & \int_p\left( 2\pi
\right)^dK'\left(
p^2\right) \left( {\delta S \over \delta \phi_p}{\delta S\over \delta
\phi_{-p}}-{\delta^2S\over \delta \phi_p\delta \phi_{-p}}\right) + d
S\nonumber \\
        & + & \int_p \left( 1-{d\over 2}+{\eta \over
2}-2p^2{K'\left( p^2\right) \over K \left( p^2\right)
}\right) \phi_p{\delta S\over \delta\phi_p}-\int_p\phi_p\, {\bf p}\cdot
{\partial '\over \partial {\bf p}}\, {\delta S\over \delta \phi_p}
        \label{n8}
\end{eqnarray}
where all quantities should be understood as dimensionless.  We have
defined $K'( z) \equiv d K (z)/ d z$
and the prime in ${\partial '\over \partial
{\bf p}}$ means that the derivative does not act on the momentum
conservation delta functions in $S$. This is our final ERG equation for
a scalar theory. It is exact, and being supplied with the initial condition
$S|_{t=0}$ defines completely the renormalization group flow of the action.

\subsection{The action}

The second step is to define the theory, i.e. to
specify the space-time on which the model is defined, its fields and its
symmetries.  Then, after characterizing the model we are instructed to
take an action which is the most general one consistent with the stated
symmetries.  In our example this would be a power-like interaction with
arbitrary number of derivatives invariant under space rotations and
discrete $Z_2$-symmetry.
The next step would be to use our equation to obtain the exact
renormalization flows of our theory.

This program is, however, as nice as unrealistic.  It is obviously
impossible to deal with such a large action.  And even in the case the
$\beta$-functions were computed somehow, the infinite set of
equations would be intractable.  Consequently, one has to develop a
reasonable approximation.

In the fermionic case we will use a mixed approximation consisting in
expansion in derivatives and truncation in the number of fields.  With
this purpose in mind we will consider an example of the mixed
approximation in the bosonic case.  This will also illustrate the
calculation of fixed points and their critical exponents within the ERG
approach.  A short review of other approximations used in the literature
and discussion of their effectiveness is given at the end of the next
subsection.  Namely, we consider a scalar field theory on the Euclidean
space with $Z_2$ symmetry and we keep all local terms up to six fields
and two derivatives.  The number of the derivatives is chosen to be the
minimum one which allows, in principle, a non-vanishing anomalous
dimension.  Therefore, we take the following initial interaction action:
\begin{eqnarray}
S_{int} & = & (2\pi)^{-d}\int_p a_1 \phi_p \phi_{-p} +
(2\pi)^{-3d} \int_{p_1,\cdots,p_4} \left(a_2+ b_2 p_1^2\right)
\phi_{p_1} \cdots \phi_{p_4}\ \delta (\Sigma p_i)\ \nonumber \\
        & + & (2\pi)^{-5d}\int_{p_1,\cdots,p_6} \left(a_3+ b_3
        p_1^2 \right)
        \phi_{p_1} \cdots \phi_{p_6}\ \delta (\Sigma p_i)  \nonumber
\end{eqnarray}
where $a_1, a_2, a_3, b_2$ and $b_3$ are real coupling constants.

\subsection{Fixed points and critical exponents}

It is a simple exercise to substitute the above
action into Eq.~(\ref{n8}) and obtain after a bit of algebra the complete
set of $\beta$-functions within this approximation.
The system of $\beta$-functions will depend on the particular shape
of the regulating
function $K(x)$ through four real quantities:
\begin{equation}
\alpha =\int_pK'\left( p^2\right) ,\ \ \ \beta =
\int_p p^2 K' \left( p^2\right) ,\ \ \ \gamma =
K'\left( 0\right) ,\ \ \ \delta =K'' \left( 0\right).  \label{pc1}
\end{equation}
However it can be proved that, within our approximation, the real
dependence on the scheme is given not by four
but only by two independent parameters. This simplification can be made
explicit after the following redefinition:
\[
a_1\to {1 \over \gamma} a_1,\ \ \
a_2\to {1\over \alpha \gamma }a_2,\ \ \ a_3\to {1\over \alpha^2\gamma
}a_3,\ \ \ b_2\to {\delta \over \alpha \gamma^2} b_2,\ \ \ b_3\to
{\delta \over \alpha^2\gamma^2} b_3 .  
\]

So, in these new variables the set of $\beta$-functions reads,
\begin{equation}
0 =\eta -y \left( 12 b_2-8 a_1^2\right)   \label{pceta}
\end{equation}
\begin{eqnarray}
&&\dot{a_1} = \left( 2+\eta \right) a_1 - 12 a_2 - 6 b_2 /x +
4 a_1^2\nonumber \\
&&\dot{a_2} =\left( 4-d+2\eta \right) a_2-30 a_3 - 10 b_3/x+
16 a_1 a_2\nonumber \\
&&\dot{a_3} =\left( 6-2d+3\eta \right) a_3+24 a_1 a_3+
16 a_2^2\nonumber \\
&&\dot{b_2} = \left( 2-d+2\eta \right) b_2 + 16 a_1 a_2+
16 a_1 b_2 - 20  b_3\nonumber \\
&&\dot{b_3} =\left( 4-2d+3\eta\right) b_3 +{192\over 5} a_2 b_2 +
24 a_1 b_3 + 24 a_1 a_3 + {144\over 5} a_2^2,  \label{pc3}
\end{eqnarray}
where $y\equiv \delta /\gamma^2 $ and $ x\equiv \alpha \gamma /\beta
\delta $.

The value of the coupling constants at fixed points
will be sensitive to $x$ and $y$.  The critical exponents in
principle are independent on them.  However, the exponents depend on
the scheme because of the truncation, as one can immediately realize
noting that the anomalous dimension $\eta $ (which in the vicinity of a
fixed point becomes one of the critical exponents) depends almost
linearly on $y$ (some non-linear dependence enters through $a_1$ and
$b_2$). A similar pattern of scheme dependence was found in 
Ref.~\cite{BHLM}.

To be concrete, let us consider briefly the numerical results that can
be obtained from Eq.~(\ref{pc3}) in $d=3$.  To this end we find first
the fixed points of the theory:  the points in the space of couplings
for which all the $\beta$-functions vanish.  That is, we solve the
following system of equations:
\begin{eqnarray}
&&0=\left(2+\eta \right) a_1-12 a_2- 6 {b_2\over x} +4 a_1^2\nonumber \\
&&0=\left(1+2\eta\right) a_2-30 a_3-10 {b_3\over x} +
     16 a_1 a_2\nonumber\\
&&0=3\eta\ a_3+24 a_1 a_3+16 a_2^2\nonumber \\
&&0=\left(-1+2\eta\right) b_2+16 a_1a_2+16a_1b_2-20 b_3\nonumber \\
&&0=\left(-2+3\eta\right) b_3+{192\over5} a_2b_2+24 a_1b_3+24a_1a_3 +
 {144\over 5} a_2^2.    \label{pc5}
\end{eqnarray}
where $\eta$ is defined in Eq.~(\ref{pceta}).

By resolving the second, third and fifth equations in
(\ref{pc5}) with respect to  $a_2$, $a_3$ and $b_3$ respectively
one can easily reduce this system to a pair of
non-linear coupled algebraic equations for $a_1$ and $b_2$.
These remaining equations have to be solved numerically.
The trivial point, where all the couplings are zero, is always
a solution and corresponds to the Gaussian fixed point.
As it is known, in the model under consideration there is
also one non-trivial fixed point, the Wilson
fixed point. The critical behavior of the model at this point
belongs to the same universality class as
the Ising model in $d=3$.
The system of equations (\ref{pc5}) has many other
 zeroes besides the two corresponding to the physical fixed points.
 The appearance of spurious
solutions is a common feature of the polynomial approximation, and as
we will discuss later, is very difficult to handle
(see Refs.~\cite{Morrisi}, \cite{HKLM}). However, analysing
successive approximations it is possible to identify the
physical solution among the fictitious ones: the latter do not
stabilize as we
increase the order of the approximation.
In our simple scalar model we can manage high orders of the
polynomial expansion,
so we can easily find the true fixed point. This is not the
case for fermionic systems
where a six-degree polynomial in fields contains more than hundred
terms within the approximation with two derivatives. We will discuss
these issues more extensively later in this article.

Having found the true fixed point, the next step is to study the
asymptotic behavior of the coupling constants flow in the
vicinity of it, governed by the critical exponents.
They can be computed by linearizing the RG transformations near
the fixed point. That is, if $g_i$ is a generic coupling constant,
then its variation in the vicinity of a fixed point $g^*=(g^*_1,
g^*_2, \ldots g^*_i, \ldots)$ is
approximated by $\delta \dot g_i=\dot g_i=R_{ij}|_{g^*}\delta g_j$,
where $g_{i}(t)=g^*_i+\delta g_{i}(t)$ and $R_{ij}$ is the matrix
$\partial \dot g_i / \partial g_j$.
The eigenvalues of $R_{ij}|_{g^*}$ can be identified with critical
exponents.
The biggest one, $\lambda_{1}$, is the inverse
of the exponent $\nu$, which governs the correlation
length in the critical domains,
and the second one $\lambda_2$, is minus the exponent
$\omega$ associated to the
slope of the $\phi^4$ operator $\beta$-function at the critical point.

For our system (\ref{pc3}) the matrix of linear deviations reads:
\[
\pmatrix{2-\eta+8 a_1&-12&0&-6/x&0\cr
16 a_2&1-2 \eta+ 16 a_1&-30&0&-10/x\cr
24 a_3&32 a_2&-3 \eta + 24 a_1&0&0\cr
16 (a_2+b_2)&16 a_1&0&-1-2\eta+16 a_1&-20\cr
24 (a_3+b_3)&{288\over 5} a_2+{192\over 5} b_2&24 a_1&
192/5 a_2&-2-3\eta+24 a_1\cr}    
\]
where $\eta$ takes the value given in the first line of Eq.~(\ref{pc3})
and all the variables are evaluated at the fixed point.

It is a simple exercise to carry out the above program in our
example.  The numerical analysis of the solutions of (\ref{pc5})
shows that the anomalous dimension $\eta$ is almost linear in
$y$ (for fixed $x$).  On the contrary, for any fixed $y$ the function
$\eta(x)$ has a maximum at some value $x^*(y)$.
Consequently we used this minimal sensitivity criterion \cite{PMS} 
to fix the parameter $x$ at $x=x^*(y)$.
Unfortunately, due to the monotonous
dependence of the solution on $y$ we are unable to set it by a similar
prescription.  A more careful analysis requires the study of the
dependence of the critical exponents on this last parameter, and,
perhaps, following this direction one could fix it by applying the
principle of minimal sensitivity.  We made this study for the critical
exponents $\nu$ and $\omega$ but with our working precision we could not
find any noticeable oscillation. This, perhaps indicates the poorness
of the approximation.

Within the range $0.1 \leq y \leq 1$ we computed the anomalous dimension
and the first two critical exponents ($\nu$ and $\omega$).
Considering the scantiness of our truncation the
results for the anomalous dimension and the critical exponents $\nu$ and
$\omega$ are quite good.  As we mentioned above the values of $\eta$ as
a function of $y$ are monotonous ranging from $0.016$ when $y=0.1$ to
$0.1$ when $y=1$.  Notice that this range includes the ``known" value of
the anomalous dimension ($\eta={0.035 \pm 0.005}$).  Within our
accuracy, the value of $\nu$ in this range is constant,
\begin{equation}
\nu = 0.53 \pm 0.015    \label{pc8}
\end{equation}
and $\omega$ ranges between
\begin{equation}
0.86 \leq \omega \leq 0.96.  \label{pc9}
\end{equation}
The known values for $\nu$ and $\omega$ are, respectively, $\nu=0.635
\pm 0.005$ and $\omega=0.8 \pm 0.05$ \cite{expc}.

Now let us discuss other approximations and compare with our results.
As it was already
mentioned in the Introduction, one of the simplest approximations is the
local potential approximation in which the interaction Lagrangian
just reduces to the
effective potential \cite{HH} (see Refs.~\cite{LPA} - \cite{Maggiore}
for further applications).  In this case the flow equation is a partial
differential equation, and the fixed point equation
is a non-linear ordinary differential equation.
Though this equation has a family of solutions only one of them
is finite for every value of the field and we immediately recognize it
as the Wilson fixed point
For this approximation the anomalous dimension is
zero and the values
of the critical exponents depend on the particular ERG equation used.
Using a sharp cutoff ERG equation for the effective potential 
the authors of Ref.~\cite{HH} obtained $\nu=0.687$, $w=0.595$. 
In Ref.~\cite{Morrisiia} these critical exponents were calculated
from the flow equation for the Legendre effective action with
a power-like additive cutoff. The results there are $\nu=0.66$,
$w=0.628$. With a Polchinski type equation
within the same approximation the authors of Ref.~\cite{BHLM} obtained
$\nu=0.649$ and $w=0.66$.

Further simplification of the local potential approximation can be
achieved by representing the effective potential as a polynomial in the
field \cite{Margaritis}, \cite{HKLM}.  In this case the differential
fixed point equation transforms into a system of the algebraic
equations, but many spurious solutions appear.  The presence of such
fictitious solutions is thus the characteristic feature of the
approximations based on truncations in the number of fields.
Analyzing successive orders of the approximation it is possible
in some cases to recognize the true fixed point solution, at least
for not very low dimensions (for example for $d > 8/3$).
Another and more serious problem is the convergence of the polynomial
approximation as the order of
the polynomial increases. Indeed there are some evidence that the
polynomial approximation does not converge,
at least for the sharp cutoff approximation \cite{Morrisi}, even though
the low orders give reasonably good numerical results for the
critical exponents.  Thus, for $d=3$ the polynomial approximation of the
Wilson fixed point with 7 terms gives $\nu=0.657$ and $w=0.705$
\cite{HKLM}.

In general better numerical results are obtained in the framework of the
derivative expansion \cite{der} - \cite{BHLM}, where the
interaction Lagrangian is expanded in powers of momenta.
Since we are interested in the low-energy properties of the model,
contributions of terms with high powers of momenta are expected to
be sub-leading. No general solid results on the convergence of
this expansion have been obtained so far (see, however, a discussion in
Ref.~\cite{Morrisiic}). The local potential
approximation is thus the leading order of the derivative expansion.
When the next-to-leading order (i.e. terms with two derivatives) is
taken into account the numerical results improve in general. Thus, for
the ERG equation of Ref.~\cite{Morrisiia} $\eta=0.054$, $\nu=0.618$ 
and $w=0.897$. For the Polchinski type equation the critical
exponents turn out to be scheme dependent, as in our example above. In
\cite{BHLM} it was shown that for a certain reasonable variation of
the scheme parameters the ranges of variation of the critical exponents
are the following:  $0.019 < \eta < 0.056$, $0.616 < \nu < 0.637$ and
$0.70 < w < 0.85$.

We see that the mixed approximation with only 6 terms, studied in
this article, works rather well comparing to other methods.  Its obvious
advantage is simplicity.  However, it possesses the drawbacks of both
the polynomial approximation and the derivative expansion. 
We expect that corrections due to higher
orders in the action should improve the results (\ref{pc8}) and
(\ref{pc9}).
In principle, they also can be improved by choosing the best suited
renormalization scheme, as it has been done in Ref.~\cite{BHLM}.

What should be learned from this example is the procedure which enables
to calculate the critical exponents starting from the expressions
for the $\beta$-functions (\ref{pc3}). The computation for
fermions within the
mixed approximation will be basically a translation of the formulas
of this section.

\section{ERG equation for fermionic theories}

In this section
we derive the ERG equation for a pure fermionic field
theory on the Euclidean space of dimension $d$.
The equation will be very similar to that for bosons.

The action is splitted, as usual,
\[
S=S_{kin}+S_{int}\ ,   
\]
where $S_{int}$ is
an arbitrary function of the fields and derivatives or momenta (if we work
in the momentum space) and $S_{kin}$ is a regulated version of the
usual kinetic term,
\[
S_{kin}=\int_p\overline\psi_{-p}P^{-1}_{\Lambda }\left( p\right)
\psi_p\ ,    
\]
where 
$P_{\Lambda }$ is now the matrix
\[
P_{\Lambda }\left( p\right) =\left( 2\pi \right)^d
{K \left( p^2 \over \Lambda^2 \right) \over p^2}i \rlap/p\ .
\]
$K \left( z\right) $ is a cutoff function with the same properties
as that for bosons.

The starting point of the derivation is now
the relation 
\begin{eqnarray}
&   & \biggl< \biggl( {\delta \over \delta \psi_p}
-\overline\psi_{-p}P^{-1}_{\Lambda } + \overline\chi_{-p}
Q^{-1}_{\Lambda }\biggr) \biggr. \dot
P^{-1}_{\Lambda }\biggl. \biggl( {\delta \over \delta
\overline\psi_{-p}}+P^{-1}_{\Lambda }\psi_p-Q^{-1}_{\Lambda
}\chi_p\biggr) \biggr>  \nonumber \\ 
& = & - \Bigl< {\rm tr} \left(
P^{-1}_{\Lambda } \dot P_{\Lambda }
\right) \delta \left( 0\right) \Bigr> - \left<
\left( \overline\psi_{-p}P^{-1}_{\Lambda }-\overline\chi_{-p}
Q^{-1}_{\Lambda }\right) \dot
P_{\Lambda
} \left( P^{-1}_{\Lambda }
\psi_p-Q^{-1}_{\Lambda }\chi_p\right) \right> ,\nonumber
\end{eqnarray}
where the trace is taken over the spinor indices. 
This is the counterpart of
Eq.~(\ref{n4}) of the previous section  
and, as there, it can be used to identify the rate of change of
the kinetic term.

Imposing the independence of the generating
functional ${\cal Z}=<1>$ on the scale $\Lambda $, we come, after a bit
of algebra, to the relation
\begin{eqnarray}
\left< \dot S_{int}\right> & = & - \int_p \left<
\eta \, \overline\psi_{-p} P^{-1}_{\Lambda }\psi_p +
\overline\psi_{-p} \dot P^{-1}_{\Lambda }\psi_p\right> - {\eta \over 2}
\int_p \left< \overline\psi_p {\delta S_{int} \over \delta \overline\psi_p}
+ \psi_p {\delta S_{int} \over \psi_p} \right> \nonumber \\
& + & \int_p\left<
{\eta \over 2}\, \overline\chi_{-p} Q^{-1}_{\Lambda }
\psi_p + {\eta \over 2}\, \overline\psi_{-p} Q^{-1}_{\Lambda }
\chi_p + \overline\chi_{-p}\dot Q^{-1}_{\Lambda
}\psi_p+\overline\psi_{-p} \dot Q^{-1}_{\Lambda }\chi_p \right> + \left<
\dot f_{\Lambda }\right> ,\nonumber
\end{eqnarray}
where the anomalous
dimension is similarly defined by 
\[
\dot\psi_p={\eta \over 2}\, \psi_p,\ \ \
\dot{\overline\psi}_p ={\eta \over 2}\, \overline\psi_p.  
\]

We can write an equation for $\left<
\dot S\right>$ by just adding the contribution of the kinetic term.
It will be satisfied if
\begin{eqnarray}
\dot S & = & \int_p\left( {\delta S\over
\delta \psi_p}\dot P_{\Lambda }\left( p\right){\delta S\over \delta
\overline\psi_{-p}}-{\delta \over \delta \psi_p}\dot P_{\Lambda }\left(
p\right) {\delta S\over \delta \overline\psi_{-p}}\right) \nonumber \\
       & + & \int_p\left[
{\delta S\over \delta \psi_p} \left( \dot P_{\Lambda }\left( p\right)
P^{-1}_{\Lambda }\left( p\right) \right) \psi_p -
\overline\psi_{-p}\left( P^{-1}_{\Lambda }\left( p\right) \dot P_{\Lambda
}\left( p\right) \right) {\delta S\over \delta
\overline\psi_{-p}}\right] .\nonumber
\end{eqnarray}

Similarly, the equations
for the terms
containing $\chi_p$ and $\overline\chi_p$ and the term 
$f_{\Lambda}$ without the sources and the fields are
\[
Q_{\Lambda }\left( p\right) =
P_{\Lambda }\left( p\right) \tilde Q\left( p^2\right) \ , \
f_{\Lambda }=
- \int dt \int_p \tilde Q^{-2}\left( p^2\right) \overline\chi_{-p}\dot
P^{-1}_{\Lambda }\left( p \right) \chi_p \ ,    
\]
with $\tilde Q\left(p^2\right) $ being a scalar
function which evolves according to the equation $\dot {\tilde
Q}\left( p^2\right) =(\eta/ 2) \tilde Q\left( p^2\right) $.

Finally, after performing the pertinent rescalings, we arrive
at the equation we will use in the sequel,
\begin{eqnarray}
\dot S & = & \int_p 2\left( 2\pi \right)^d K' \left(
p^2 \right) \left( {\delta S\over
\delta \psi_p}i \rlap/p{\delta S\over \delta
\overline\psi_{-p}} -{\delta \over \delta \psi_p}i\rlap/p
{\delta S\over \delta \overline\psi_{-p}}\right)     \label{feq} \\
       & + & d S
+ \int_p \left( {1-d+\eta \over 2}-2p^2{K' \left( p^2
\right) \over K \left( p^2\right) }\right) \left(
\overline\psi_p{\delta S\over \delta \overline\psi_p}+\psi_p {\delta S\over
\delta \psi_p}\right) - \int_p \left( \overline\psi_p p^{\mu }
{\partial '\over \partial  p^{\mu }}{\delta S\over \delta
\overline\psi_p}+\psi_p p^{\mu }{\partial '\over \partial p^{\mu }}
{\delta S\over \delta \psi_p}\right) .\nonumber
\end{eqnarray}

Notice that, once the above ERG equation is derived,
it would be easy to obtain a similar one
for a model involving also other fields and interactions
just by combining the
manipulations here with those of
the previous section.
(The resemblance of Eq.~(\ref{feq})\ 
and the Polchinski type equation for scalar
theories is pretty evident.)

As a final comment about our equation let us comment that we present 
it on Euclidean
space as it is customary in the field. For our purposes, however, there is
nothing special about the Euclidean formulation, as finally what one obtains
is just a set of relations among coupling constants.  In fact,
we have also derived the counterpart of Eq.~(\ref{feq})\ for
Minkowski space. It is not so nice because of the presence of an extra 
imaginary unit coming from the functional derivatives of the Minkowskian
``Boltzmann" factor $e^{i{\cal S}}$ in the second term.  Nevertheless,
with this equation we have computed the $\beta$-functions for a
simplified action (one without operators with six fields) in much the
same way we will explain later for Euclidean space: they are finite,
real and consistent with the desired symmetries, as they should be.
We have not proceeded further, but the parallelism between them and their
Euclidean counterparts strongly
supports the common lore
that both should contain the same physical
information and that the choice of space is much a matter of taste.
Nevertheless, it would probably be nice to afford a complete 
calculation in Minkowski space.

\section{The fermionic action}

We turn now to consider spin $1/2$ fields.  We want to apply
the machinery to a concrete model, which we define in this section
through its symmetries, then discuss sensible
truncations and, finally, comment on the prescriptions which we have 
actually used to build its action in a systematic way.

We choose a general theory with $N$ spin-$1/2$ fields on the two-dimensional
Euclidean space which obeys the discrete symmetries of 
parity, charge conjugation and, to obtain reflection positive 
Green functions, reflection
hermiticity (see Ref.~\cite{book} for a precise definition of them) 
in addition to the
invariance under the standard Euclidean transformations. 
Further we will impose
the invariance under the chiral symmetry transformations of 
$U(N)_R\times U(N)_L$.

The next step is to choose an appropriate truncation. A reasonable idea 
is to use a kind of the derivative expansion, which was proved to be 
quite efficient in bosonic theories \cite{der} - \cite{BHLM}. However, 
unlike in the scalar case, the approximation without derivatives 
(effective potential) does not work and the leading order 
should include terms both without derivatives and with 
one derivative. 
The reason is simple: 
Eq.~(\ref{feq})\ contains, due to the sum over polarizations, the factor 
$\rlap /p$ 
in the Kadanoff terms, while a similar equation for bosons does not (see,
for instance, Eq.~(\ref{n8})).
Another significant difference is that in the scalar case a general
potential contains an infinite number of independent
operators (powers of fields), whereas for finite $N$
a general function of fermionic variables with fixed number of
derivatives has in any case a finite number of terms due to the 
Grassmann nature of the field.

However, for a given but large enough $N$ (the case which we will
be mainly interested in) the number of different structures 
grows fast as the number of derivatives
increases. In practice, the approximation is intractable already in the 
order with $3$ derivatives unless the degree of the 
polynomial of the fields is also restricted.
We consider two approximations here. The first one is the
pure derivative expansion in next-to-leading order (with all terms
without derivatives and with one derivative). The second
approximation is the mixed one where the number of
derivatives and {\it also} the power of fields are truncated. 
In this calculation we consider terms with up to three derivatives
and up to six fermionic fields. This allows for non-vanishing
anomalous dimension.

The final preparatory step is to write down the action. To construct it
systematically we consider each symmetry in turn
and derive the restrictions it imposes. 
For the second approximation,
we will neither comment the details of the construction, nor write down
the whole action, which in this case consists of $107$
independent operators.
The interested reader 
is referred to \cite{CKM} for a thorough discussion. 
Here as an example we present only terms ${\cal S}^{(s,q)}$ ($s$
stands for the number of fermionic fields and $q$ for the number of
derivatives) of the action with 2 and 4 fields. In the momentum
representation they read:
\begin{eqnarray}
{\cal S}^{(2,1)} & = & p_{12}^{-\,j}iV_{12}^j, \nonumber \\
{\cal S}^{(4,0)} & = & g_1(S_{12}S_{34}-P_{12}P_{34})+g_2V_{12}^j
                       V_{34}^j, \label{action40} \\
{\cal S}^{(4,2)} & = & \{ m_1p_{12}^{+\,2}+m_2p_{12}^-\cdot p_{34}^-
+m_3p_{12}^{-\,2}\}\times (S_{12}S_{34}-P_{12}P_{34})  \nonumber \\
      & + & \{r_1p_{12}^{+\,2}+r_2p_{12}^-\cdot p_{34}^-
     + r_3p_{12}^{-\,2}\}\times V_{12}^jV_{34}^j  
     + \{s_1p_{34}^{+\,j}p_{12}^{+\,k}+s_2p_{12}^{-\,j}p_{12}^{-\,k}
+s_3p_{12}^{-\,j}p_{34}^{-\,k} \nonumber \\
     & + & s_4p_{34}^{-\,j}p_{12}^{-\,k}\}\times V_{12}^j V_{34}^k
+tp_{34}^{+\,j}p_{34}^{-\,k}\epsilon^{jk}(S_{12}S_{34}-P_{12}P_{34}),
          \label{action42}
\end{eqnarray}    
where we introduced the following notations:
\begin{equation}
S_{12}\equiv
\overline\psi^a\left(p_1\right)\psi^a\left(p_2\right),\
\ \
P_{12}\equiv \overline\psi^a\left(p_1\right)\gamma_s\psi^a\left(p_2\right),\
\ \
V_{12}^{j}\equiv \overline\psi^a\left(p_1\right)\gamma^j\psi^a\left(p_2\right)
        \label{genoper}
\end{equation}
(the flavour indices are summed up) and $p_{kl}^{\pm\,j}\equiv
(p_k\pm p_l)^j$. ${\cal S}^{(2,1)}$ is of course the kinetic term. 

In the first case, however, the action is written as an infinite
series in increasing powers of
derivatives with coefficients being arbitrary functions of scalar operators 
without derivatives built out of the fermions. 
We work in the $N\to \infty$ limit.

{}From the discussion above we see that there are two independent 
scalar operators without derivatives, which in the coordinate
representation are (compare with Eq.~(\ref{genoper}))
\begin{equation}
 R(x) = \overline\psi^{a}(x) \gamma^{j} \psi^{a}(x)  
        \overline\psi^{b}(x) \gamma^{j} \psi^{b}(x)  \label{roper}
\end{equation}
and
\[
 U(x) = \overline\psi^{a}(x) \psi^{a}(x) \overline\psi^{b}(x) \psi^{b}(x) -
        \overline\psi^{a}(x) \gamma_{s} \psi^{a}(x) 
	\overline\psi^{b}(x) \gamma_{s} \psi^{b}(x) .  
\]
Then, the most general action in our approximation can be written as 
\begin{eqnarray}
   S &= & \int d^{2}x \left\{ A(R(x),U(x)) \right. \nonumber \\
   & + & \overline\psi (x) \gamma^{j} \partial_{j} \psi (x) B(R(x),U(x))
                                       \nonumber \\
   & + & \overline\psi (x) \gamma^{j} \psi (x) 
     \overline\psi (x) \gamma^{i} \partial_{j} \psi (x) 
     \overline\psi (x) \gamma^{i} \psi (x) C(R(x),U(x)) \label{actone} \\
   & + & \overline\psi (x) \gamma^{j} \psi (x) 
   (\overline\psi^{a}(x) \partial_{j} \psi^{a}(x)
   \overline\psi^{b}(x) \psi^{b}(x) -
    \overline\psi^{a}(x) \gamma_{s} \partial_{j} \psi^{a}(x) 
	\overline\psi^{b}(x) \gamma_{s} \psi^{b}(x)) D(R(x),U(x)) \nonumber \\
   & + & \left. \overline\psi (x) \gamma^{j} \psi (x) 
       \epsilon^{ij} (\overline\psi^{a}(x) \partial_{i} \psi^{a}(x) 
	\overline\psi^{b}(x) \gamma_{s} \psi^{b}(x) - 
	\overline\psi^{a}(x) \psi^{a}(x) 
	\overline\psi^{b}(x) \gamma_{s} \partial_{i} \psi^{b}(x)) 
	E(R(x),U(x)\right\}, \nonumber 
\end{eqnarray}
where $A$, $B$, $C$, $D$ and $E$ are arbitrary functions of the 
operators $R$ and $U$.
The standard kinetic term is contained in the second term of this
action. Thus, the operator $B$ should be normalized by the condition 
$B(0,0)=1$. 

\section{Pure derivative expansion for fermions}

In this section we will study fixed point solutions of the fermionic 
ERG equation in the next-to-leading order of the derivative expansion. 
The critical exponents will be analysed elsewhere. 
To derive the flow equations we write the action (\ref{actone})\ in the 
momentum representation, 
understanding the operator functions $A$, $B$, etc. as being 
expanded formally in powers of $R$ and $U$, and then 
substitute it into the ERG equation (\ref{feq}). After long but 
straightforward calculation a system of equations for generating functions 
$A(r,u)$, $B(r,u)$, $C(r,u)$, $D(r,u)$ and $E(r,u)$ can be obtained. 
These functions are the same as their operator prototypes introduced 
above, but with the operator arguments $R(x)$ and $U(x)$ being replaced by 
the real c-numbers $r$ and $u$ respectively. The equations for the 
generating functions are the following:
\begin{eqnarray}
\dot{A}(r,u) & = & 2 A - 2(1-\eta) (r\partial_{r}+u\partial_{u}) A 
   +  4 \alpha N B - 4 \alpha (r\partial_{r}+u\partial_{u}) B
                  \nonumber \\ 
    & + & 2 \alpha ( N rC - r C - r^{2} \partial _{r}C - 
    ru \partial_{u} C) 
  - 2\alpha [ (r+u) D + ru ( \partial _{r} D + \partial _{u} D) ];
          \nonumber \\
\dot{B}(r,u) & = & \eta B - 2(1-\eta) (r\partial_{r}+u\partial_{u}) B 
       + 2 \gamma r (\partial_{r} 
       A)^{2} - 2 \gamma u (\partial _{u} A)^{2}; \nonumber \\
\dot{C}(r,u) & = & -(2-3\eta) C - 2(1-\eta)  (r\partial_{r}+u\partial_{u}) C - 
    4 \gamma (\partial_{r} A)^{2}; \nonumber \\
\dot{D}(r,u) & = & -(2-3\eta) D - 2(1-\eta) (r\partial_{r}+u\partial_{u}) D - 
    4 \gamma (\partial_{r} A)(\partial _{u} A); \nonumber \\
\dot{E}(r,u) & = & -(2-3\eta) E - 2(1-\eta) (r\partial_{r}+u\partial_{u}) E + 
    2 \gamma (\partial_{u} A)^{2} + 4 \gamma (\partial_{r} A) 
    (\partial _{u} A), \label{eqabcde}
\end{eqnarray}
where $\alpha$ and $\gamma$ are 
scheme parameters defined by Eq.~(\ref{pc1}). As before the dot 
means the derivative with respect to the flow parameter $t$. 
We see that the function $E(r,u)$ appears only in the 
last equation and is completely determined by the function $A(r,u)$.

There is only one solution regular in $r$ and $u$ 
at $r=u=0$. It is $u$-independent and is given by the following 
expressions for the generating functions:
\begin{equation}
A(r,u) = - 2\alpha N + g r, \; \; \; \;  
B(r,u) =  1 + \gamma g^{2} r, 
C(r,u) =  - 2 \gamma g^{2}, \; \; \; 
D(r,u)  =  0,  \; \; \; E(r,u)=0, \label{solabcde}
\end{equation}
where the normalization of the kinetic term has been taken 
into account. The anomalous dimension $\eta=0$ for this fixed point.  
Note that Eq.~(\ref{solabcde})\ is valid for any $N$. 
This solution is polynomial in $r$ 
and is characterized by the arbitrary continuous 
parameter $g$. Since $r$ corresponds to the operator (\ref{roper})\ in the 
original action, the term $ g r$ in the expression for $A(r,u)$ 
corresponds to the operator $g V_{12}^j V_{34}^j$, which gives 
rise to the $U(1)$ Thirring like term. This agrees with the known fact that 
in the chiral Gross-Neveu model the sector, corresponding to this term, 
decouples from the rest of the model and is at the fixed point 
for any value of $g$. However, the coupling constant, corresponding 
to the $SU(N)$ sector, is zero for the solution (\ref{solabcde}). Because 
of the too low order of our approximation, we do not think that 
this fixed point can be reliably identified with any of the known 
fixed points of the Gross-Neveu model. 

Let us mention for completeness that the system (\ref{eqabcde})\ has 
another formal fixed point solution. 
The anomalous dimension $\eta=0$ again, and the generating functions 
do not depend on $r$. The expression for $A$ is given by 
\[
 A(u) = A_{0} + G_{0} \int_{u_{0}}^{u} {d u' \over
        1-2G_{0} \alpha \gamma N \ln (u'/u_{0})} , 
\]
where $A_{0}$, $G_{0}$ and $u_{0}$ are integration constants (only 
two of them are independent). One can 
see that derivatives of $A$ are not regular at $u=0$. Other 
generating functions for this solution are expressed through 
similar integrals. It is not clear for us whether such formal solution 
has any physical meaning. 

\section{Mixed approximation for fermions}

In this section we will consider the approximation including the terms with 
up to three derivatives and up to six field operators. Simplification due to 
the truncation in the number of fields allows us to increase the number 
of the derivatives in the approximation, that in its turn amounts 
to non-zero anomalous dimension, as we will explain shortly. 
Having constructed the corresponding action, as it was explained 
above, the next step 
is to substitute it into the ERG equation (\ref{feq})\ and to compute the
$\beta$-functions or flow equations 
of our model within the chosen approximation.

One comment is in order here. In the actual calculation 
of the $\beta$-functions we used an extended action which was larger
than the one discussed in the previous section 
in order to have some extra check of the equations.
Namely, we considered an action written in a basis of functionals 
with two fermions and one and three derivatives, four fermions with 
zero and two derivatives and six fermions with one and three derivatives 
which were restricted by $U(N)$ symmetry, parity and Euclidean
invariance only (that is, we imposed neither reflection
hermiticity, nor charge conjugation, nor chirality).
We then projected the space of functionals generated by this basis 
onto the subspace of functionals invariant under all  
symmetries of the theory and its direct complement. 
The required flow equations were obtained after we performed the first
projection, while the projection onto the 
complementary subspace provided us a consistency
check of the calculation. The latter defined a set of null equations, or 
identities, which had
to be satisfied along the renormalization flow. Indeed, since the
initial action at $t=0$ is symmetric, appearance of any non-zero 
non-symmetric term at $t>0$ would indicate an anomaly, 
which, as we have argued, does not appear. 

\subsection{Generalities}

The next step is to find the fixed point solutions, that
is, the sets of values of the coupling constants which make all 
the $\beta$-functions to vanish.
They correspond to stationary points of the RG flows, thus providing us
with the first information of how the phase diagram of the system looks like.

$\dot S=0$ is equivalent to the system of 106 non-linear algebraic
equations. A simplification comes from the observation that 
all the coupling constants
of operators with six fields must enter linearly. The reason is that the
only source of non-linearity in Eq.~(\ref{feq})\ is the first 
term of its r.h.s.,
and neither ${\cal S}^{(6,1)}$ nor ${\cal S}^{(6,3)}$ contribute to this 
term within our approximation.
Resolving the subsystems $\dot{\cal S}^{(6,1)}=0$ and 
$\dot{\cal S}^{(6,3)}=0$ 
with respect to the couplings of six-field operators we reduce 
the system to a set of only 13 non-linear equations, plus an equation for
the anomalous dimension $\eta$ that satisfies 
\begin{equation}
\eta = 4 \alpha [-m_1 + m_2 + m_3 + s_1 + s_2 + s_3 + s_4 + t - 2 N ( r_2 +
     2 s_3 + s_4 )].  \label{eqeta}
\end{equation}
Here $N$ is the number of flavours;
$\alpha $, $\beta $, $\gamma $, $\delta $ are
scheme dependent parameters defined by the same formulas as (\ref{pc1}); 
the rest of coefficients of (\ref{eqeta})\ are different coupling constants
of the four-field functionals of our action, see Eq.~(\ref{action42}). 

The appearance of the above dependence just reflects the freedom in 
choosing the renormalization scheme, or the cutoff function in our case.
Furthermore, although the $\beta $-functions contain four
scheme parameters, we will see that, as in the scalar field case,
the fixed point solution and critical exponents depend
only on two combinations of them. 

We found the fixed point numerically for finite $N$, whereas
for $N\to \infty$ we could solve it analytically in the 
leading order of the $1/N$ expansion. Results of both 
calculations are presented below. 

After the fixed points are found,
the behaviour of the theory near each of them is controlled by the
critical exponents. One of them is already known once we solved our system 
of equations:
it is the anomalous dimension at the fixed point.
The rest are found by linearizing the ERG flow equations near a
chosen fixed point $g^{*}$ in the same way as it was explained for the
bosonic case in Sect.~2.3. The eigenvalues of $R_{ij}|_{g^{*}} = 
\left(\partial \dot g_i / \partial g_j \right) |_{g^{*}}$ 
are identified with the critical exponents. 
They can be viewed as the anomalous dimensions 
of the corresponding operators in the vicinity of the fixed point. 

Finally, let us turn again to the issue of scheme dependence. As we have 
already mentioned the values of the coupling constants
at a fixed point are scheme dependent, thus reflecting that they are not
universal quantities.
Critical exponents, on the other hand, are universal and 
should be scheme independent.  Nonetheless,
due to
the truncation, the scheme dependence will inevitably
appear. We will proceed in a common way \cite{BHLM}\ and try 
to find a scheme where the effect of the dependence will be the least. 
To this end,
we will apply the principle of minimal
sensitivity in order to fix the scheme dependence. 

\subsection{$N\to \infty$}

Now we are going to define 
the large N expansion in our model which
allows to obtain analytical results in the leading order in $1/N$. 
To this end we substitute each coupling
constant $g_i$ by $N^{z_i}g_i$
and study the limit $N\to \infty $ keeping $g_i $ fixed.
$z_i$ can be any real
number, but for the sake of simplicity we restrict ourselves to 
integer values.

We are looking for the sets $\left\{ z_i\right\} $ for which 
the $\beta $-functions for redefined couplings are finite and, 
if possible, non-trivial in the $N \to \infty$ limit. 
With the above requirements, we found two different patterns of 
the $1/N$ expansion, which lead to inequivalent results. We will label
them by I and II, and discuss in turn.

The Type I solution is obtained by taking $z_i=-1$, where $i$ runs over
all four-fermion couplings. With this definition, the
anomalous dimension vanishes in leading order in $1/N$ and at the
fixed point the coupling constant $g_1$ of the operator
$(S_{12}S_{34}-P_{12}P_{34})$ in ${\cal S}^{(4,0)}$, 
Eq.~(\ref{action40}), is fixed,
\[
g_1   = -1/( \alpha  \gamma)  
\]
whereas the coupling $g_2$ of the operator $V_{12}^j V_{34}^j$,
giving rise to the $U(1)$ Thirring excitations, is arbitrary.

The characteristic
polynomial $P(\lambda )$, associated to the matrix 
$R_{ij}|_{g^{*}}$, can be computed analytically:
\begin{eqnarray}
P(\lambda) & = & \lambda^2\ (\lambda+2)^{12}\ (\lambda +4)^{83}\ 
        (\lambda+6)\ (\lambda^2+ 6 \lambda -8)      \nonumber  \\
&\times & \left(-\lambda^5 - 12 \lambda^4+
(8 w-44) \lambda^3 +(64 w-16) \lambda^2+(32 w +64)\lambda-(128 w 
+256)\right), \nonumber
\end{eqnarray}
where $w=\beta \delta /(\alpha \gamma)$ and $z=\delta /\gamma^2$.
The critical exponents can be read from $P(\lambda)$. There are $100$
scheme independent eigenvalues, most of them
coinciding with the canonical values $0, -2, -4$ and $-6$.  The
non-trivial ones are $-3+\sqrt{17}=1.1231...$ and 
$-3-\sqrt{17}=-7.1231...$. The rest of the eigenvalues are given by 
the roots of the polynomial
\[
Q(\lambda)=
-\lambda^5-12 \lambda^4+(8 w-44) \lambda^3+(64 w-16) \lambda^2+
(32w +64)\lambda-(128 w +256) 
\]
and are $w$-dependent.
If $w<0$, that includes, for instance, the case of the exponential cutoff
function $K(z)=e^{-\kappa z}$, the most
relevant critical exponent is $\lambda_1= 1.1231...$

The fixed point solution contains two free continuous
parameters $g_2$ and $m_{1}$. 
This is the
expected result for the chiral Gross-Neveu type model because 
$U(1)$ Thirring like excitations (which in our action are controlled by
$g_2$) decouple from the rest and this subsystem is conformal
invariant (i.e.~it is at the fixed point) for any value of $g_2$.  For the
$SU(N)$ part there exists a discrete set of fixed points,
the fixed point of Dashen and Frishman being one of them.
It is reached when the constant $g_1$
is of the order $1/N$, as in our case.  So we can expect that 
our solution is the fixed point of Ref.~\cite{DF}.  However, the
values of the anomalous dimension do not match. For the
cited fixed point it is
non-vanishing in leading order in $1/ N$ and not zero as we have found.
This discrepancy with the exact result of Dashen and Frishman could be
caused by the low accuracy of our truncation. 
We cannot reject, however, the possibility
of having found
a different fixed point (see Ref.~\cite{ofp}).

For the Type II solution we let a combination of coupling constants
to be of order 1, instead of $O(N^{-1})$.  We do not know if this 
possibility is an accident of the truncation or if it would 
survive in a complete calculation.
The most significant feature of this solution
is its non-zero anomalous dimension,
\[
\eta={4\over 3} - {\beta \gamma\over 6\alpha}=
  {4\over 3}-{1\over 6}{w \over z}. 
\]

Unlike the previous case, we could not find the exact analytical
expression for the characteristic polynomial. However, by computing
numerically the eigenvalues for different values of $z$ and $w$ we could
deduce some exact results. It turned out that none of the 
critical exponents coincide with
their canonical counterparts. Moreover, most of them are functions of
the combination ${w\over z}$. Thus, there are 82 eigenvalues
$\lambda=-{w\over 2z}$, 8 eigenvalues equal to ${2\over 3}(1-{w\over
2z})$ and 4 equal to $2-{w\over 2z}$.  The remaining ones are 
functions with more complicated dependence on $w$ and $z$ 
then the ratio $w/z$ (and even a few have 
a non-vanishing imaginary part,
which is not unusual in approximations based on truncations).
We studied numerically
the most relevant critical exponent, which turned out to belong to the
class with non-simple dependence on $w$ and $z$,
for a wide range of values of the scheme parameters. 
As it also happens in the
scalar case, for any value of $z$ this exponent always has a minimum
at some $w=w^*$. Such behaviour suggests to use the
minimal sensitivity criterion to fix the parameter $w$ to its critical
value $w^*$.  Unfortunately, due to the monotonous dependence of the
exponent on the parameter $z$ in the range analysed, we were unable to
fix it by a similar prescription.  We show in Table 1 
$\lambda_1^{*}(z)=\lambda_1(w^*,z)$ and $w^{*}$ for 
some values of $z$ and
the corresponding anomalous dimension.

\begin{table}
\centerline{
\begin{tabular}{|c|l|l|l|l|}                     \hline
                & $z=0.1$ & $z=0.5$ & $z=1.0$ & $z=2.0$ \\ \hline
 $\lambda^*_1$  & $2.258$ & $2.239$ & $2.217$ & $2.175$ \\ \hline
 $w^{*}$        & $0.122$ & $0.616$ & $1.250$ & $2.610$ \\ \hline
 $\eta$         & $1.130$ & $1.128$ & $1.125$ & $1.116$ \\ \hline
\end{tabular}}
\caption{ Local minimum $\lambda_{1}^{*}$ 
of $\lambda_1(w,z)$, the most relevant critical exponent,
for different values of $z$. $w^*$ is the value of $w$ at which the minimum 
is reached and $\eta$ is the corresponding value of the anomalous dimension.}

\end{table}

\subsection{Finite $N$}

For finite number of flavours we could not find fixed point solutions 
analitically, so we studied the zeroes of the $\beta$-functions
numerically. In general the
number of different solutions of a system of coupled non-linear equations
is not known a priori, and common programs for
root-finding (such as the FindRoot routine of
{\it Mathematica}) do not give all its solutions.
Nevertheless, after some experience was acquired and relying on the 
results for high $N$ we could choose a reasonable range of values 
of the coupling constants and examine it minutely in searching for the 
fixed point solutions. 

A more serious problem 
is to discriminate between zeros, which correspond to 
real fixed point solution, and spurious roots, which are artefacts 
of the truncation.
This problem, which appears in the bosonic
case too, is
perhaps the Achilles' heel of the approximations based 
on truncations \cite{Morrisi}, \cite{HKLM}. We present the class of solutions 
of which we are more confident.
These are mainly the ones which asymptotically match with some solution
clearly identified in the framework of the large $N$ expansion.

To begin with, let us 
select a particular scheme and find the solution for
different values of $N$. We take 
$w=-2$ and $z=0.5$,
corresponding to the exponential regulating function $K(x)=e^{-x^2}$.
We will analyse the dependence on these parameters later on.

There is a sequence of fixed point solutions for various (interger) $N$
which asymptotically approaches the type I solution of the
$N \to \infty$ limit discussed above. For this sequence  $N\eta$ 
increases with $N$ and tends to $4.87...$, while
the most relevant critical exponent $\lambda_1$ decreases with $N$
and asymptotically approaches the value $1.1231...$, in a full 
agreement with our previous results on the $1/N$ expansion.  
The second eigenvalue was found to be complex, that is
apparently an artefact of the approximation used. 
Another mismatch is that the type of the solutions of the sequence does not
fit the type of the $N \to \infty$ fixed point. As we wrote above, 
the latter is the two-parameter continuous line of solutions parametrized by 
$g_{2}$, which is related to the Thirring $U(1)$ sector, and $m_{1}$,
whereas the solutions for finite $N$ are isolated. Most likely, this 
qualitative change of the type of the solution occurs when we go from
infinite $N$ to, though large, but finite $N$ within our
approximation. However, we have not studied this phenomenon in detail.
We present in Fig. 1 
the plots of $N \eta$ and $\lambda_1$ as 
functions of $N$.

\begin{figure}[thb]
\vspace{-4cm}
\centerline{\psfig{figure=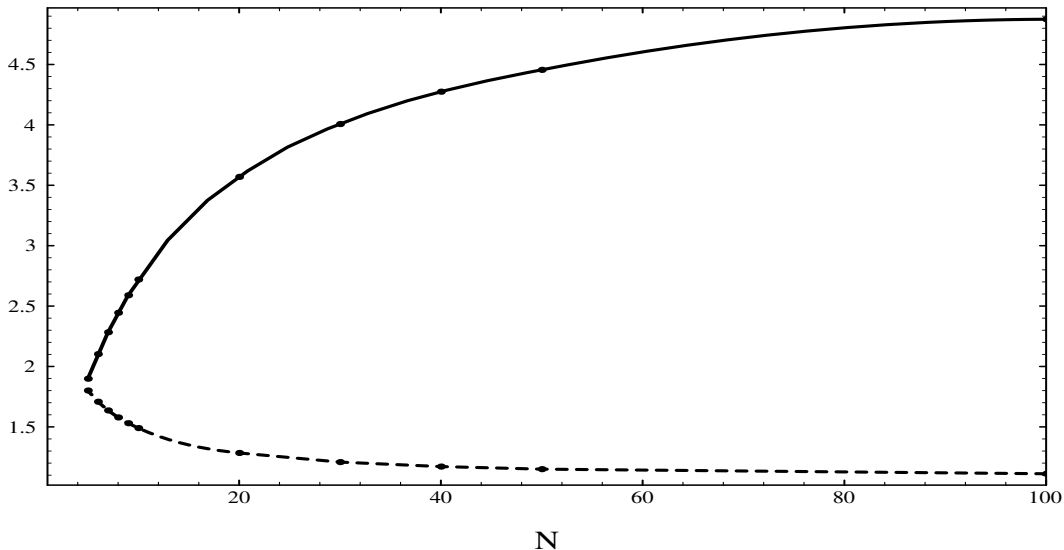,height=15cm,width=14cm}}
\vspace{-4cm}
\caption{ $N\cdot\eta$ (solid line) and $\lambda_1$
(dashed line) as functions of $N$. The corresponding fixed point 
solution matches the Type I solution of the large $N$ limit.}
\end{figure}

Let us discuss now the dependence of the solutions on
the renormalization scheme. The parameter $z$ enters the flow 
equations only through the anomalous dimension and, for this reason, 
the dependence of the fixed points solutions on $z$ is rather
simple: for instance, it is almost linear for $\eta$. The dependence
on $w$ is more complicated, and we studied the behaviour of $\eta$
and $\lambda_{1}$ under the change of $w$ for fixed $N$ and $z$. 
In this analysis, as in the scalar case, we looked for some non-linear
$w$-dependence so that we could invoke the principle
of minimal sensitivity to fix the value of this parameter and reduce 
the scheme dependence.
To this end, we take $z=0.5$ and $N=1000$.
The curve $\eta$ vs.~$w$ is monotonous and decreases with $w$, while
the first eigenvalue $\lambda_1$ reaches its minimum value
$\lambda_1=1.12511$ at $w=-45$, which increases as we
lower $N$: it is equal to $1.1273$ for $N=500$ (and the minimum is 
reached at $w=-23$), $1.146$ for $N=200$ (at $w=-10$), $1.1519$ for
$N=100$ (at $w=-8$), $1.695$ for $N=10$ (at $w=-2.4$) and finally,
$2.560$ for $N=3$ (at $w=-0.5$).  

For a sequence of fixed point solutions which match the Type II
solution the situation is different. 
For $N$ large enough,
(say $N \geq 1000$), the numerical solutions are in good agreement 
with the infinite $N$ analytical result
(for example the value of $\eta=1.99$ for $z=0.5$ and $w=-2$, this is
to be compared with the exact result $\eta=2$ for $N\to \infty$).
As we lower
$N$ the values of the anomalous dimension $\eta$ and the most relevant
eigenvalue $\lambda_1$ decrease and near 
$N=142$ (actually at $N=142.8$ if we let $N$ to take non-integer
values) the solution ceases to exist and joins another branch of
fixed point solutions. 
For fixed points belonging to the second branch
$\eta$ and $\lambda_1$ remain finite as $N\to \infty$.
However, we found, that some couplings did not behave as powers 
of $N$ in this limit and,
therefore, this solution cannot be associated with a fixed point in the 
$N \to \infty$ limit 
in the sense stated previously.  At the joining point
$\eta= 1.88$ and $\lambda_1= 5.80$.  We show in Fig. 2 
the curves $\eta(N)$ and $\lambda_1(N)$.
Further studies should be carried out in order to
understand the structure of these solutions better. 

\begin{figure}[thb]
\vspace{-4cm}
\centerline{\psfig{figure=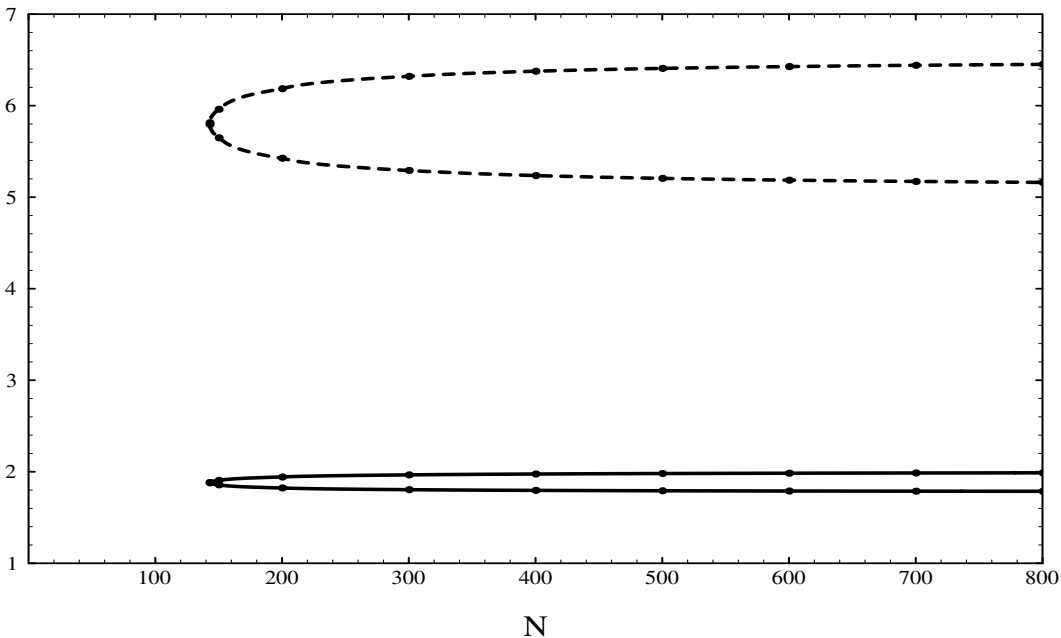,height=17cm,width=14cm}}
\vspace{-4cm}
\caption{ $\eta$ (solid line) and $\lambda_1$ (dashed line) as 
functions of $N$ for $z=0.5$ and $w=-2$.  In both
curves the upper branch corresponds to the solution which matches the
Type II solution of the large $N$ limit.} 
\end{figure}

There are also other solutions, which were found for low $N$.
For some of them either $\eta$ or
$\lambda_1$ have minima in $w$, 
but in other cases both curves are monotonous in $w$. They also
display various behaviours as $N\to \infty$. 

Finally we will consider the special case $N=1$.
Fierz transformations relate $U(N)$-covariant local operators (like
$\overline\psi^a\left(p_1\right)\psi^b\left(p_2\right)$) to 
$U(N)$ scalars (like
$\overline\psi^a\left(p_1\right)\psi^a\left(p_2\right)$). So, for $N=1$ 
the Fierz transformations establish additional relations 
between the available
operators, which give
relations
between the coupling constants that allow us to reduce the
system of the flow equations considerably.

After a bit of algebra and
discarding the trivial solution we finally get an equation for
$\eta$, 
\begin{eqnarray}
0 & = & -120 w^2  z + 288 w z^2  + \eta (13 w^2  - 132 w z +
210 w^2  z + 288 z^2  - 720 w z^2 )  \nonumber \\
& + & \eta^2  (99 w z - 90 w^2  z - 432 z^2  + 594 w z ^2) +
 \eta^3  (162 z^2  - 162 w z^2 ) .  \nonumber
\end{eqnarray}
As in the previous analysis we had to
choose some particular scheme, i.e.~fix $w$ and $z$, and solve
the equation numerically. Unfortunately, 
the system of fixed point equations is not underdetermined and
there is no room for a free parameter in the fixed point
solution as it is the case in the Thirring model. 
In fact such property takes place 
in the previous order approximation (terms with less than
three derivatives), however $\eta$ vanishes identically there. This is in a 
complete agreement with the first solution of the pure derivative expansion 
considered in Sect.~5. The reason why this
property is lost in the approximation with three derivatives is unclear for
us. Probably, one has to go to the next
order of the approximation to understand this issue. 

We solved the equation numerically for different values of $w$
and $z$. As in previous examples the fixed point solutions are almost linear
in the parameter $z$, their behaviour with $w$ is more complicated. 
However, in the range of values studied,
we did not find any non-monotonous dependence on $w$ either of the critical
couplings or of the anomalous dimension.
We present in Table 2 some of the results for $\eta$.

\begin{table}
\centerline{
\begin{tabular}{|l|l|l|l|l|}                     \hline
          & $=-0.1$ & $w=-0.5$ & $w=-1.0$ & $w=-2.0$ \\ \hline
$z=0.1$   & $1.763$ & $3.691$  & $6.316$  & $11.747$ \\ \hline
$z=0.5$   & $1.418$ & $1.790$  & $1.418$  & $3.388$  \\ \hline
$z=1.0$   & $1.376$ & $1.559$  & $1.811$  & $2.349$  \\ \hline
$z=2.0$   & $1.354$ & $1.445$  & $1.569$  & $1.834$  \\ \hline
\end{tabular}}
\caption{ Values of $\eta$ for $N=1$ for different values of the 
scheme parameters $z$ and $w$.}
\end{table}

\section{Discussion and conclusions}

In this article we gave an account of main ideas and techniques of the
ERG approach. We presented steps of the derivation of the ERG
equation, which plays the crucial role, for scalar and fermionic
theories. We discussed main approximations, which have been developed
so far, for solving it and showed how some characteristics of models,
like fixed points and critical exponents, can be computed. We also reviewed
briefly some non-perturbative results, obtained within this approach
for the scalar $Z_{2}$-symmetric model in three dimensions and for 
the two-dimensional Gross-Neveu type model. In the latter case the
results are quite recent. As we argue above, within the $1/N$ expansion
the Type I fixed point solution is an excellent candidate for
the Dashen-Frishman fixed point, whereas the other one presents
evidences to be a new fixed point, with quite intricate properties,
not discussed previously in the literature.

We saw that in scalar theories truncation in the power of fields (like
in the polynomial approximation of the local effective potential or in
the mixed approximation) leads to numerous spurious solutions with the
subsequent problem of identification of the true physical one, especially
when the order of the polynomial is high enough. However, if the
pure derivative expansion is used, the spurious solutions and the problem
of identification do not appear. We demonstrated that in the case of
fermionic theories the situation is similar. One should only keep in mind
that for finite $N$, due to the Grassmannian nature of the fields,
terms with $l$ derivatives do not contain more than $k_{max}(l) = 4N +
2[l/2]$ spinors (here $[x]$ stands for the integer part of $x$).
Thus, each term of this expansion is a polynomial in fermionic fields.
When we use the mixed approximation, i.e. we restrict the action
to the number of fields less than $k_{max}(l)$ for a given $l$, as it
was the case in most of the examples considered in Sect.~6, we observe
the appearance of many spurious solutions. When we carry out the
pure derivative expansion, either in a general form (see Sect.~5) or
by taking into account all possible operators with powers till
$k_{max}(l)$ in fields, like in the example with $N=1$ in Sect.~6.3
($l \leq 3$ there), we see that no spurious solutions appear. Thus, for
$N=1$ there are only three fixed point solutions, besides the trivial one,
two of them having complex coupling constants, thus being rejected at once.
It would be interesting to perform a pure derivative expansion for,
say, $N=2$ and check if the above features hold.

All approximations which include derivatives contain scheme parameters,
which cannot be eliminated by redefinition of the coupling constants.
Such scheme dependence does not appear, however, in the lowest
approximations without derivatives. This is clearly illustrated
by our scalar example in Sect.~2.3. For the approximation without
derivatives, i.e. when $b_{2}$ and $b_{3}$ are zero, $\eta=0$ and
the equation (\ref{pceta}) is not present, no scheme parameters appear
in the first three equations of the system (\ref{pc3}). When the terms
with two derivatives are taken into consideration, the fixed
point solution and critical exponents depend on the scheme parameters
$x$ and $y$. This dependence on the regulating (cutoff) function is
analogous to the renormalization scheme dependence in the perturbation
theory calculations. For observable quantities, like critical exponents,
this dependence is unphysical and is an artefact of the approximation.
We used the principal of minimal sensitivity to "minimize" their
dependence on the scheme.

An interesting feature is the seemingly good results
for the large $N$ limit in the fermionic model.
At the computational level this is related to the fact
that at the leading order of the $1/N$ expansion the system
of the flow equations simplifies dramatically
and no room is left for spurious solutions. The same simplification
occurs in scalar models as well \cite{WH}, \cite{Tetradis}.
It would be useful to understand deeper reasons for this.
Apparently, improvement of results in the large $N$ limit is a quite
general feature of the ERG approach.

Recent developments of the ERG formalism allow us to conclude that
it can be regarded as a reliable non-perturbative method of studies
of physical models. Calculational techniques, developed in
scalar theories, were proved to be effective enough in calculations
of the numerical characteristics of the models and analysis of
renormalization flows. In fermionic theories much less work has been
done so far. It would be interesting to probe similar calculational
techniques in higher approximations, than discussed here, or in
other physical models. For example, the ERG formalism should be
extended to higher dimensions. The equation (\ref{feq}) is prepared for that.
The number of spinor structures will increase in this case and,
therefore, one will have to handle more terms in the action.
However, we guess that,
due to the greater complexity, two derivatives may be sufficient to
obtain interesting results, or at least, according to the standard rule
that quantum effects become less important when the dimension is
increased, we hope that the results will be more transparent 
within the same approximation.

We did not discuss gauge theories in this article.  This is because of
lacking of a satisfactory ERG formalism for the moment.  The main
problem is to maintain the gauge invariance of the ERG equation and the
renormalization flow, which is obviously broken when cutoff functions of
the type (\ref{n3}) are introduced in order to set the physical scale.
Development of an effective formalism for gauge theories seems to be the
most challenging problem in the ERG approach nowadays.

\vspace{4mm}
{\bf Acknowledgements} 
\nobreak

We are sincerely indebted
to J.I.~Latorre, who encouraged us in this work and provided
us fruitful suggestions during its completion.
Discussions with A.A.~Andrianov and D.~Espriu are also acknowledged.
Yu.~K.~and E.~M.~thank the Department ECM of the University of 
Barcelona for warm
hospitality and friendly atmosphere during their stay there.
Yu.~K.~also would like to thank the Department of Theoretical Physics
of the UAM for hospitality. 
This work was supported in part by funds provided
by CICYT under contract AEN95-0590 and by the M.E.C. grants
SAB94-0087 and SAB95-0224.

\end{document}